\documentclass[conference]{IEEEtran}
\IEEEoverridecommandlockouts

\usepackage{cite}
\usepackage{amsmath,amssymb,amsfonts}
\usepackage{algorithmic}
\usepackage{graphicx}
\usepackage{textcomp}
\usepackage{xcolor}
\usepackage{makecell}

\usepackage{siunitx}
\usepackage{hyperref}
\usepackage[nolist]{acronym}
\usepackage[flushleft]{threeparttable}
\usepackage{placeins}
\usepackage{svg}                                   
\usepackage[percent]{overpic}

\usepackage{mathtools}
\usepackage{booktabs}
\usepackage[nolist]{acronym}
\usepackage{pbalance} 
\usepackage[export]{adjustbox}

\usepackage{eso-pic}    

\def\BibTeX{{\rm B\kern-.05em{\sc i\kern-.025em b}\kern-.08em
    T\kern-.1667em\lower.7ex\hbox{E}\kern-.125emX}}
\newcommand{\revise}[1]{{\textcolor{black}{#1}}}

\newcommand{\RNum}[1]{\uppercase\expandafter{\romannumeral #1\relax}}   

\DeclareSIUnit\db{dB}                           
\DeclareSIUnit\dbi{dBi}                         
\DeclareSIUnit\dbm{dBm}                         
\DeclareSIUnit\watthour{Wh}                     
\DeclareSIUnit\mbps{Mbps}                       
\DeclareSIUnit\kbps{kbps}                       
\DeclareSIUnit\bps{bps}                         
\DeclareSIUnit\msInference{ms/inference}        

\begin{document}
\bstctlcite{IEEEexample:BSTcontrol} 

\AddToShipoutPictureBG*{
  \AtPageUpperLeft{%
    \put(0,-40){\raisebox{15pt}{\makebox[\paperwidth]{\begin{minipage}{21cm}\centering
      \textcolor{gray}{This work has been submitted to the IEEE for possible publication. Copyright may be transferred without notice, after which this version may no longer be accessible.
      } 
    \end{minipage}}}}%
  }
  \AtPageLowerLeft{%
    \raisebox{25pt}{\makebox[\paperwidth]{\begin{minipage}{21cm}\centering
      \textcolor{gray}{This work has been submitted to the IEEE for possible publication. Copyright may be transferred without notice, after which this version may no longer be accessible.
      }
    \end{minipage}}}%
  }
}

\title{BodySense: An Expandable and Wearable-Sized Wireless Evaluation Platform for Human Body Communication}

\author{
\IEEEauthorblockN{
Lukas Schulthess, 
Philipp Mayer, 
Christian Vogt, 
Luca Benini,
Michele Magno
}\\

\IEEEauthorblockA{Dept. of Information Technology and Electrical Engineering, ETH Z{\"u}rich, Switzerland}
}

\maketitle
\IEEEpeerreviewmaketitle

\begin{abstract}
\revise{Wearable, wirelessly connected sensors have become a common part of daily life and have the potential to play a pivotal role in shaping the future of personalized healthcare. A key challenge in this evolution is designing long-lasting and unobtrusive devices.
These design requirements inherently demand smaller batteries, inevitably increasing the need for energy-sensitive wireless communication interfaces.
Capacitive \ac{HBC} is a promising, power-efficient alternative to traditional RF-based communication, enabling point-to-multipoint data and energy exchange.
However, as this concept relies on capacitive coupling to the surrounding area, it is naturally influenced by uncontrollable environmental factors, making testing with classical setups particularly challenging.}

\revise{This work presents a customizable, wearable-sized, wireless evaluation platform for capacitive \ac{HBC}, designed to enable realistic evaluation of wearable-to-wearable applications.
Comparative measurements of channel gains were conducted using classical grid-connected- and wireless \ac{DAQ} across various transmission distances within the frequency range of $\mathbf{4\mskip3mu}$MHz to $\mathbf{64\mskip3mu}$MHz and
revealed an average overestimation of $\mathbf{18.15\mskip3mu}$dB over all investigated distances in the classical setup.}

\end{abstract}
\vspace{10pt}

\section{Introduction}
Wearable, intelligent, and unobtrusive sensor nodes monitoring the human body and its environment have attracted significant attention from researchers and industry \cite{svertoka_2020_industry_wearables, design_wearabe_ates_2022}. These devices, equipped with wireless interfaces and an increasing number of sensors, are already a commercial reality for sports \& fitness, with expanding capabilities aimed at collecting valuable data for human-centric healthcare \cite{flex_electronics_wu_2023}. Advances in system-on-chip integration and flexible electronics have enhanced sensor integration on dynamic, non-planar surfaces like the human body, enabling localized physiological signal acquisition and processing \cite{roadmap_flex_luo_2023, flex_wrist_pulse_lee_2023, jose_2024_e_textile}. However, distributing multiple sensors on body locations with beneficial sensing characteristics inevitably increases the stress on the communication interface due to growing synchronization and connection overhead \cite{iob_survey_celik_2022, eff_sensor_node_adway_2024}. 

\begin{figure}[!t]
    \centering
    \begin{overpic}[width=1\columnwidth]{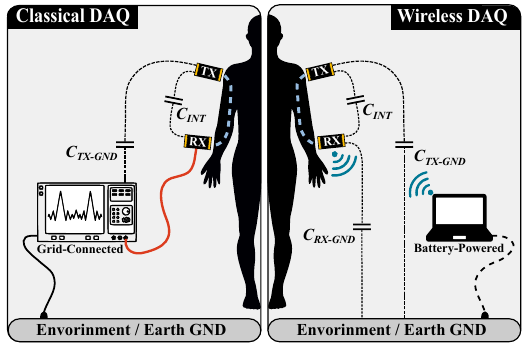}
    \end{overpic}
    \caption{\revise{Comparison between \ac{HBC} \ac{DAQ} setups:
    (left) Classical \ac{DAQ} setup using grid-powered measurement tools, (right) Miniaturized and fully-wireless \ac{DAQ} setup}.
    }
    \vspace{-5mm}
    \label{fig:hbc_highlevel}
\end{figure}

Today's wearable devices rely predominantly on \ac{RF} wireless transmission, already one of the most expensive subsystems, consuming significant power and limiting device lifetime and usability \cite{ WBAN_energy_basic_2016, tinybird_schulthess_2023}. Due to its widespread adoption and interoperability, \ac{BLE} has emerged as the de facto standard for wearable communications. 
However, \ac{BLE}, like other low-power wireless \ac{RF} transmission technologies, may not be the optimal choice for all wearable applications. It faces challenges such as relatively high power consumption and vulnerability to security risks \cite{wi-R_datta_2023}.

These limitations can make \ac{BLE} less suited for ensuring consistently reliable, energy-efficient, and pervasive operation in wearables.
To reduce the impact of the communication interface on the energy budget, \ac{RF} devices can be periodically deactivated, a method known as duty cycling. However, while duty cycling can be effective, it inevitably introduces latency and does not eliminate power consumption during idle listening. An alternative approach is the design of always-on wireless receivers that have the ability to detect wireless messages of interest while consuming power in the micro to nano-watt range \cite{wur_piyare_2017, wur_shellhammer_2023}.
However, these solutions fail to enable battery-free and, thus, truly pervasive operation. A paradigm shift in communication and power management is essential to realize the vision of a \ac{WBAN} with distributed, battery-free, on-body wearables.

Emerging techniques like \acs{HBC} offer a promising alternative. By using the conductive properties of the human body as a communication medium, HBC can address the limitations of \ac{BLE} and other RF wireless communication by significantly reducing power consumption, mitigating external interference, and enhancing security against external attacks. \cite{das_body_eqs_2019, datta_2021, chatterjee_2023_annual_review}. 
Unlike conventional \ac{RF} transmission techniques, such as \ac{BLE}, \ac{UWB}, or \ac{RFID}, capacitive \ac{HBC} employs the electrical conductivity of the human body to exchange information between devices \cite{hbc_zimmerman_1996}. Thanks to that property, capacitive \ac{HBC} has the potential for secure \cite{nath_2020, yang_2022}, body-constraint, and efficient data and energy transmission \cite{shukla_2019, wpt_dong_2021, wpt_modak_2022}, making it an ideal solution for future wearable devices \cite{chatterjee_2023_annual_review}.

This work presents a wearable-sized evaluation platform for \ac{HBC}. Custom-designed and compact, it enables \ac{HBC} evaluations in wearable-to-wearable application scenarios by minimizing measurement errors caused by artificially strengthened return paths. Comparable in size to commercial smartwatches, the modular evaluation platform, named \textit{BodySense}, is tailored to allow testing under realistic conditions. This platform will be used to evaluate the potential for creating energy-efficient body sensor networks. In particular, this article presents the following contributions.

\begin{enumerate}
\item The design of a versatile, expandable, wearable-sized wireless evaluation platform for \ac{HBC}, enabling practical and reliable measurements. 
\item An analysis of the channel gain across multiple transmission distances at frequencies between \qty{4} {\mega\hertz} and \qty{64}{\mega\hertz}, for both wearable-to-wearable and wearable-to-grid-connected application scenarios.
\item The demonstration of the significance of a fully wearable evaluation setup for capacitive \ac{HBC} by comparing the channel gains between classical and wearable \ac{DAQ}.
\end{enumerate}

The rest of this article is organized as follows: \revise{Section II} positions this work within the context of the \revise{state of the art}. \revise{Section III} details the design and implementation of the \textit{BodySense} hardware platform. The evaluation setup topology is outlined in \autoref{sec:experimental_setup}, with the corresponding measurement results presented in \autoref{sec:results}. Finally, \autoref{sec:conclusion} summarizes the key findings and concludes this work.

\section{Background and Related Work}\label{sec:related_work}

\begin{figure}[!t]
    \centering
    \begin{overpic}[width=1\columnwidth]{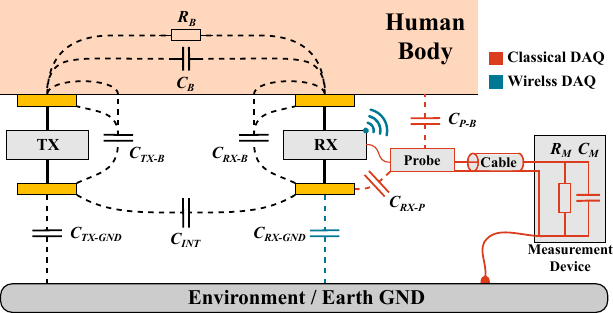}
    \end{overpic}
    \caption{\revise{Equivalent circuit for classical DAQ (red) and wireless DAQ (blue).
    }}
    \vspace{-5mm}
    \label{fig:hbc_equivalent}
\end{figure}

Based on the underlying coupling principle, \ac{HBC} is divided into three main methods: \textit{Capacitive coupling (CC)}, \textit{galvanic coupling (GC)}, and \textit{magnetic coupling (MC)}.
Capacitive \ac{HBC} uses a signal electrode connected to the body and a floating ground electrode. It achieves low forward path loss throughout the body, but its performance depends on the return path and load capacitances \cite{m_hbc_nath_2022}, limiting its use in implantable devices \cite{9128564}.
\textit{Galvanic} \ac{HBC} injects differential signals using two body-attached electrodes. 
Signal quality predominantly depends on the electrode size, spacing, and parasitic capacitances.
\textit{Magnetic} \ac{HBC} benefits from low tissue permeability, making it effective for \ac{nLoS} communication and intra-bidy communication \cite{m_hbc_wen_2022}. However, its range is limited by the magnetic near-field.

When comparing the coupling principles presented above, capacitive \ac{HBC} shows the best performance when it comes to longer-distance and low-power body-centered communication \cite{m_hbc_nath_2022}. 
The transmitter couples a signal over the skin-attached electrode to the conductive layers under the human skin, utilizing a low-impedance forward path between the transmitters and the receivers' skin electrodes \cite{a_bora_2020}. 
The floating ground electrodes form the return path by direct capacitive coupling between each other for short distances or over the environment for longer distances. Whereas the forward path is subject to minor fluctuations caused by the coupling quality of the skin-electrode and the individual physical composition \cite{maity_modelling_2019}, the return path is heavily dependent on the environment, acting as a limiting factor for reception quality.
Inter-body coupling to other people \cite{nath_2020} or coupling to any larger conductive surface such as measurement equipment, laptop, and cables strengthens the return path, leading to over-optimistic results, see \autoref{fig:hbc_highlevel}, left. This makes it difficult to reliably quantify influences on the communication link in a controlled manner for wearable-to-wearable scenarios \cite{avlani_2020}. 
Hence, the analysis in a wearable-to-wearable scenario is necessary to eliminate overestimation in amplitude and power reception of the received signal that arises from artificially enhanced ground coupling over the attached test equipment, see \autoref{fig:hbc_equivalent}.

\begin{figure*}[!t]
    \centering
    \begin{overpic}[width=\textwidth]{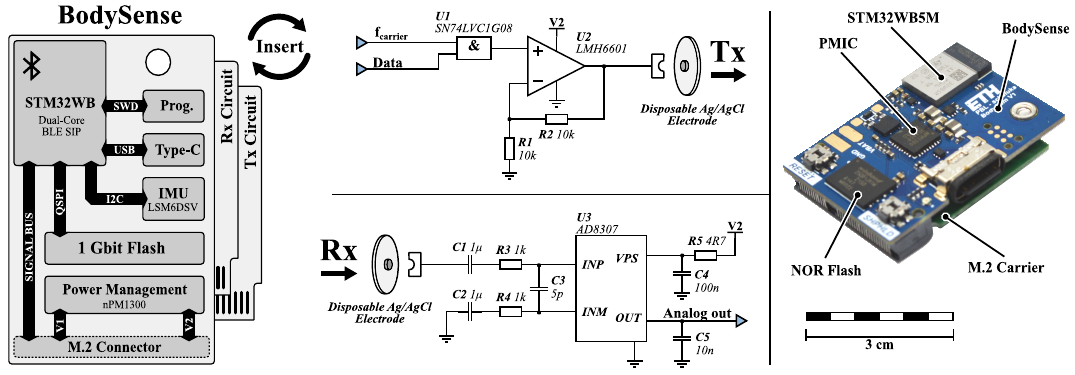}
        \put(1,33){(a)}
        \put(31,33){(b)}
        \put(31,14.5){(c)}
        \put(73,33){(d)}
    \end{overpic}
    \vspace{-6mm}
    \caption{System overview: (a) High-level block diagram of the main platform \textit{BodySense}, 
    (b) Tx carrier circuit, (c) Rx carrier circuit, (d) \textit{BodySense} platform, equipped with M.2 carrier.}
    \vspace{-5mm}
    \label{fig:system_blockdiagram}
\end{figure*}
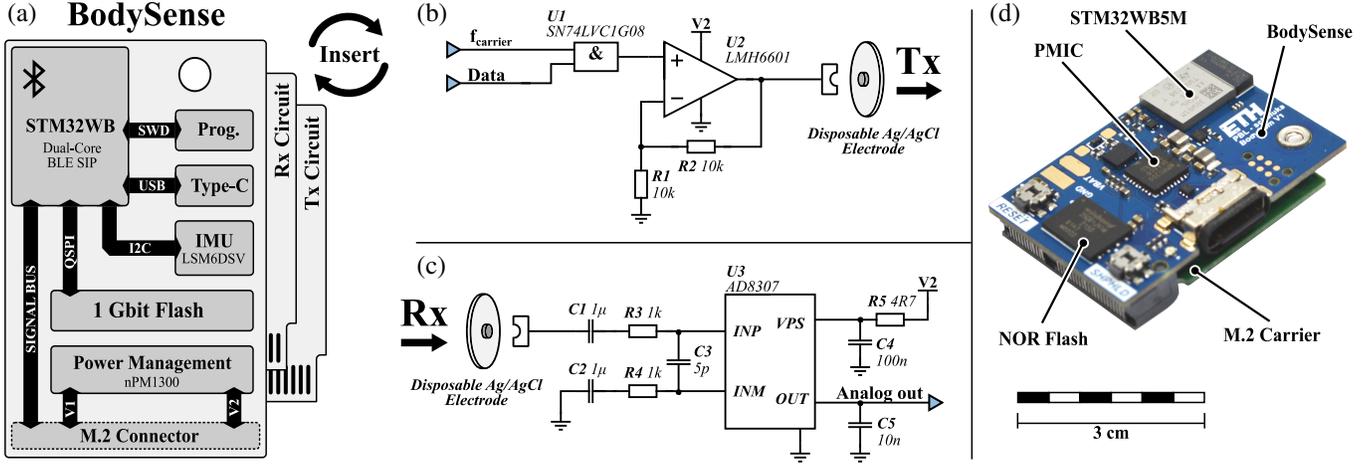

Several recent works propose chip-integrated front-ends for capacitive \ac{HBC} achieving \unit{\pico\joule\per\bit} communication links with data rates of up to \qty{20}{\kilo\bit\per\second} \cite{wpt_modak_2022, chbc_comm_maity_2019}, demonstrating a significant reduction in energy consumption for data transfer \cite{chatterjee_2023_annual_review}. 
Likewise, recent works exploit capacitive \ac{HBC} for power transmission, demonstrating the possibility of transferring up to tens of \unit{\micro\watt} over the whole body. 
Modak et al. presented a \ac{HBC} \ac{WPT} IC in a \qty{65}{\nano\meter} process, achieving power transfer of \qty{240}{\micro\watt} in a machine-to-machine configuration (close distance, perfect coupling to earth-ground) and \qty{5}{\micro\watt} in a wearable-to-wearable configuration (long distance, bad coupling to earth-ground) \cite{wpt_modak_2022}.

Although these numbers are remarkable, all works clearly highlight the challenges of capacitive \ac{HBC} due to the variability of electrical characteristics within and between humans and its reliance on an earth-ground coupled return path. These factors make direct comparisons between studies particularly difficult, as results are only partially reproducible \cite{datta_2021}.
Furthermore, previous research predominantly evaluates system performance using bulky, well-coupled laboratory equipment, often leading to overly optimistic outcomes.
In contrast, this work focuses on designing a reliable setup capable of accurately measuring \ac{HBC} performance under realistic, application-oriented conditions.

\section{System Design}\label{sec:methods}
\textit{BodySense} is an expandable and wearable-sized wireless evaluation platform for human body communication that has been designed to assess \ac{HBC} under realistic conditions.
With an overall size of \qty{33.5}{\milli\meter}~\(\times\)~\qty{22}{\milli\meter}, it is comparable to commercial smartwatches and is thus representable for wearable-to-wearable application scenarios.
\autoref{fig:system_blockdiagram} shows a simplified block diagram of the main board, detailed schematics of the designed Tx and Rx frontends, and the hardware realization.

\subsection{System Overview}
As the core platform, \textit{BodySense} integrates essential circuits, including the computational unit and power management subsystem, and provides flexible expandability via an M.2 connector. It is built around the \textit{STM32WB5MMG} module, an ultra-low-power, \ac{BLE}-enabled \ac{SiP} that integrates an ARM Cortex-M4 MCU running at up to \qty{64}{\mega\hertz} as the primary core and an ARM Cortex-M0+ hosting the \ac{BLE} stack. Additionally, the module comprises passive components for power supply stabilization, antenna matching, and clock generation. The extensive set of analog and digital peripherals, combined with its compact form factor of \qty{11}{\milli\meter}~\(\times\)~\qty{7.3}{\milli\meter} and integrated wireless connectivity, makes the module an excellent foundation for orchestrating \textit{BodySense}.

The \ac{PMIC} \textit{nPM1300} supplies power to \textit{BodySense} and its extension board while also monitoring the battery's state of charge and managing recharging. Two integrated software-configurable DC-DC converters power the mixed-signal circuitry.
For the presented Rx and Tx circuits, these converters are configured to output \textit{V1} = \qty{2.7}{\volt} and \textit{V2} = \qty{3.3}{\volt}, respectively. Load switches enable power gating of the extension board, thereby extending battery life. Additionally, a \qty{1}{\giga\bit} \textit{W25Q01JV} NOR flash provides onboard data logging capabilities.

An M.2 connector provides electrical and mechanical connectivity between \textit{BodySense} and the carrier board, offering a versatile interface for future circuit designs and expansion.
Dedicated instances of UART, SPI, and I2C peripherals are directly routed to the M.2 connector.
Additionally, \ac{ADC} inputs, timer inputs, conventional \acs{GPIO} pins, and wake-up pins are made available through the M.2 connector, providing flexibility for custom front-end designs and supporting a wide range of potential extensions.

\subsection{Application-specific M.2 carrier boards:}
Two carrier boards, a receiver (Rx) and a transmitter (Tx), were designed to demonstrate the versatility of the system and to evaluate capacitive \ac{HBC} in realistic application scenarios.
Their detailed schematics are shown in \autoref{fig:system_blockdiagram} (b) and (c).
\revise{Development-accompanying SPICE simulations, followed by verification through benchtop measurements, confirmed the correct functionality under the defined conditions.
However, variances in environmental coupling as well as in the skin impedance can vary by a factor of 10 \cite{maity_modelling_2019} and thus have a much higher impact on the measurements than the frontend's performance variation caused by component tolerances.
As a consequence, testing on the human body is the only way to obtain realistic results.}

\textit{Rx Circuit:} Disposable Ag/AgCl electrodes are used to capture the transmitted signal from the body.
\revise{A passive band-pass filter with corner frequencies f\textsubscript{L}~=~\qty{160}{\hertz} and f\textsubscript{H}~=~\qty{70}{\mega\hertz} suppresses the dominant \qty{50}{\hertz} mains noise and other unwanted external signals from the environment that couple to the body. The subsequent logarithmic amplifier of type \textit{AD8307}, with an operating range from DC to \qty{500}{\mega\hertz} and a dynamic range of \qty{92}{\db}, generates an analog output voltage proportional to the received signal level on a logarithmic scale. It is stabilized using a capacitor and then fed to the \ac{ADC} on the main board for quantization. This frontend topology provides the needed high dynamic range to measure the \ac{RSS}.}
\textit{Tx Circuit:} 
The internal \ac{PLL} of the \textit{STM32WB5MMG} module is used to generate a rectangular carrier frequency f\textsubscript{carrier}, ranging from \qty{4}{\mega\hertz} up to \qty{64}{\mega\hertz}, directly accessible on the modules GPIO.
An AND gate of type \textit{SN74LVC1G08} is used in combination with an operational amplifier of type \textit{LMH6601}, featuring a high slew rate of \qty{260}{\volt/\micro\second} and a \ac{GBP} of \qty{250}{\mega\hertz}, to generate an OOK-modulated transmit signal with a low output impedance.
Finally, the signal is routed over a snap button to a disposable Ag/AgCl electrode to establish good skin contact.

\section{Experimental Setup}\label{sec:experimental_setup}

\begin{figure}
    \centering
    \begin{overpic}[width=\columnwidth]{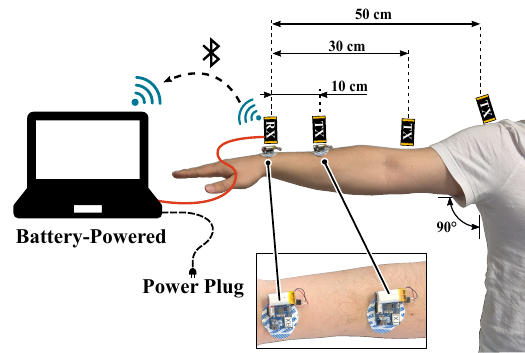}
        \put(28,55){(a)}
        \put(36,38){(b)}
    \end{overpic}
    \caption{Overview of the experimental setup for (a) wireless and (b) classical data readout. Transmitter and receiver placement are identical for both scenarios.}
    \vspace{-5mm}
    \label{fig:experimental_setup}
\end{figure}

To \revise{quantify} the effect of artificially enhanced capacitive coupling for classical \ac{DAQ} and compare it against wireless \ac{DAQ}, three measurement series for each of the two scenarios have been conducted.
One \textit{BodySense} system has been equipped with an Rx carrier placed on the test subject's upper wrist. 
An adhesive Ag/AgCl wet-gel electrode from \textit{TIGA-MED} with a diameter of \qty{48}{\milli\meter} acts as skin-electrode, ensuring a low resistive contact to the body and mechanically holds the system at its position.
The system ground plane and the battery act together as a floating electrode, closing the return path over the environment and earth-ground.
A second \textit{BodySense} system with a Tx carrier extension has been sequentially placed at distances of \qty{10}{\centi\meter}, \qty{30}{\centi\meter}, and \qty{50}{\centi\meter} to the receiver, utilizing the same Ag/AgCl wet-gel electrode as skin-electrode.
\autoref{fig:experimental_setup} provides an overview of the transmitter and receiver positions during the experiments.
Distances below \qty{10}{\centi\meter} have not been investigated as the inter-device coupling significantly strengthens the return path, making it comparable to the forward path loss \cite{yang_2022}.
For each distance, a frequency sweep from \qty{4}{\mega\hertz} up to \qty{64}{\mega\hertz} has been performed.
All measurements have been conducted in a standing position in a laboratory environment while being at least \qty{1}{\meter} away from walls and other equipment.
The arm was outstretched to the side, forming an angle of \qty{90}{\degree} between the arm and the torso's side. 
During the data collection, the subject stood still and kept the arm in a constant position to minimize movement-induced fluctuations in the environmental-coupled return path.

In the classical scenario, the receiver was directly connected over a USB Type-C cable, forwarding the collected data over USB to the plugged-in computer, see \autoref{fig:experimental_setup} (a).
For the wireless scenario, the USB Type-C cable has been removed, and the computer has been unplugged from the grid. The data is wirelessly forwarded via \ac{BLE} to the computer, as depicted in \autoref{fig:experimental_setup} (b).
Finally, the channel gain is calculated to measure the quality of the capacitive \ac{HBC} communication channel, and the received power has been extracted to set capacitive \ac{HBC} into context to conventional \ac{RF} solutions.

It is essential to carefully consider the effects of signal transmission through the human body and respect its safety limits in \ac{HBC} systems. The \ac{ICNIRP} sets limits for non-ionizing radiation exposure to the human body, defining different dosimetric quantities depending on the frequency. In the frequency range between \qty{1}{\hertz} and \qty{100}{\kilo\hertz} the current density is used as limiting metric \cite{icnirp_low} and above \qty{100}{\kilo\hertz} the \ac{SAR}-value \cite{icnirp_high}.
Further, the IEEE standard C95.1-2005 defines safety levels on human exposure to \ac{RF} electromagnetic fields \cite{c95.1-2005}.
By limiting the transmit power to maximal \qty{5}{\dbm} for our experiments, we ensured to meet the above-mentioned safety standards.
\section{Results}\label{sec:results}
\begin{figure}
    \centering
    \begin{overpic}[width=\columnwidth]{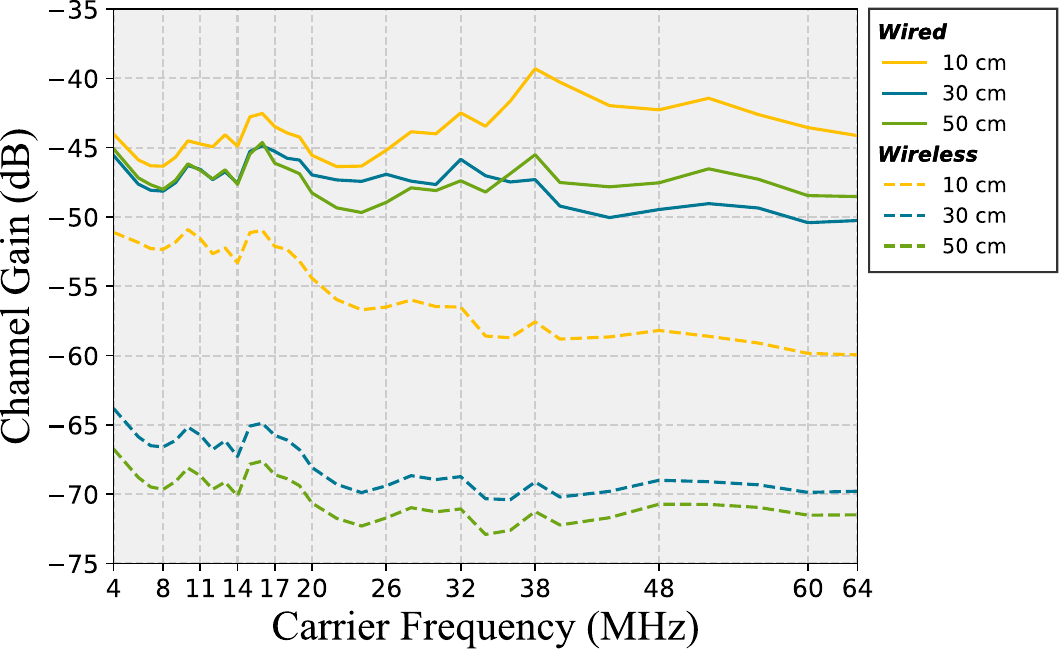}
    \end{overpic}
    \vspace{-5mm}
    \caption{Channel gains measured over different frequencies for the classical and the wireless receiver scenario. 
    }
    \label{fig:results_channel_gains}
    \vspace{-3mm}
\end{figure}

The measurement results for the classical as well as for the wireless \ac{DAQ} scenario are discussed. 
In the classical scenario (\autoref{fig:results_channel_gains}, solid lines), the lower frequencies between \qty{4}{\mega\hertz} and \qty{20}{\mega\hertz}, the channel gains show an almost identical frequency dependency for all distances. 
Whereas the more considerable distances (\qty{30}{\centi\meter} and \qty{50}{\centi\meter}) achieve almost identical results, it shows an improvement of at least \qty{2.6}{\db} for the \qty{10}{\centi\meter}-distance in this frequency region.
For higher frequencies above \qty{20}{\mega\hertz}, the difference between \qty{10}{\centi\meter} and the larger distances start to increase.
Direct coupling between the transmitters and receivers floating ground electrode, illustrated as \(C_{INT}\) in \autoref{fig:hbc_equivalent}, enhances the return path for short distances.
The largest channel gain with a peak value of \qty{-39,3}{\db} has been achieved for \qty{38}{\mega\hertz} at \qty{10}{\centi\meter} distance.
Above \qty{36}{\mega\hertz}, the \qty{50}{\centi\meter}-distance shows a superior channel gain than for \qty{30}{\centi\meter}, indicating an artificially improved return path caused by the wired data transmission.
Overall, all three measurement series of the classical scenario are close to each other, showing a maximum difference of \qty{8}{\db} at \qty{36}{\mega\hertz}.

For the results of the wireless scenario (\autoref{fig:results_channel_gains}, dashed lines),
it is interesting to see that the channel gain curves show a high correlation of 0.88 between the \qty{10}{\centi\meter} and \qty{50}{\centi\meter}, 0.95 between \qty{10}{\centi\meter} and \qty{30}{\centi\meter}, and 0.98 between the \qty{30}{\centi\meter} and \qty{50}{\centi\meter} curves. 
Similar to the results of the classical scenario, the channel gain for \qty{10}{\centi\meter} outperforms the more considerable distances by at least \qty{10}{\db}. 

\begin{figure}[t]
    \centering
    \begin{overpic}[width=\columnwidth]{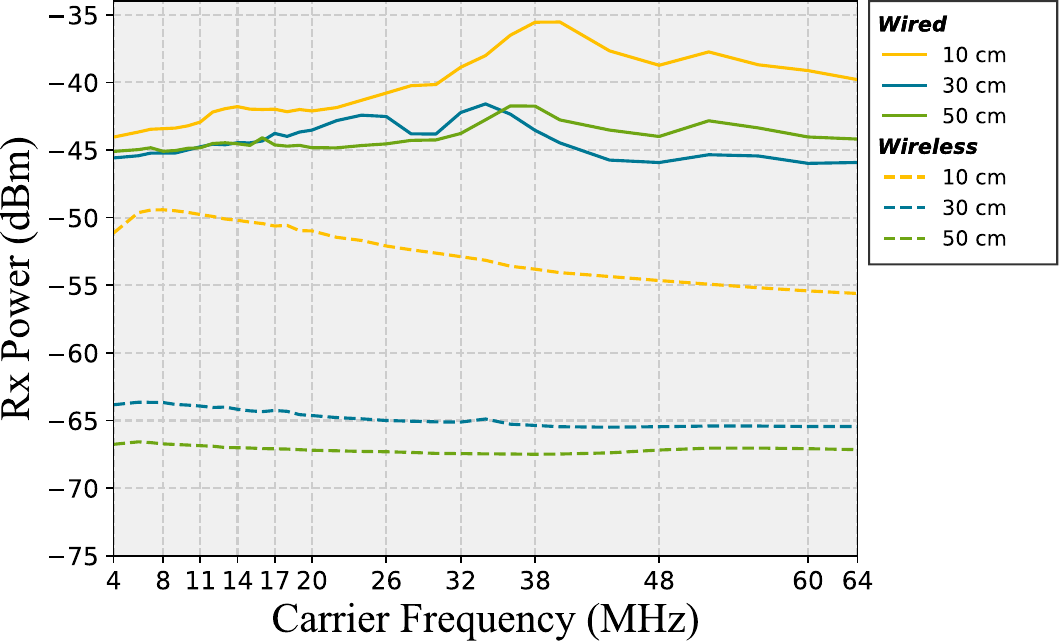}
    \end{overpic}
    \vspace{-5mm}
    \caption{Received power measured over different frequencies for the classical and the wireless receiver scenario.
    }
    \label{fig:results_rx_power}
        \vspace{-6mm}
\end{figure}

\autoref{fig:results_rx_power} shows the received power for both the classical and the wireless receiver scenario.
The wired scenario demonstrates superior performance compared to the wireless setup across all distances, achieving a maximum power reception of \qty{-35.53}{\dbm} at \qty{38}{\mega\hertz} and a distance of \qty{10}{\centi\meter}.
The Rx power behaves almost constant for longer distances in the wireless scenario, achieving a minimal power reception \qty{-67.49}{\dbm} with an overall fluctuation of \qty{0.91}{\db} over the whole frequency range.

The experimental results visualized in \autoref{fig:results_channel_gains} and \autoref{fig:results_rx_power} show that the classical scenario achieves a significantly higher channel gain than the wearable scenario.
It is essential to recognize that the \qty{10}{\centi\meter} distance outperforms the longer distances in both scenarios. 
Whereas the \qty{10}{\centi\meter}-case is only \qty{2.6}{\db} superior in the classical case, the channel gains curves for \qty{30}{\centi\meter} and \qty{50}{\centi\meter} lie at least \qty{10}{\db} below the \qty{10}{\centi\meter}-curve in the wireless setup.
This is due to the strong inter-device coupling between the transmitter and receiver at close distances.
In contrast to the wireless case, this effect is significantly less pronounced in the wired scenario, as all distances benefit from improved coupling.
When comparing the channel gain curves for the two setups, one can readily observe that the channel gain has increased by an average of \qty{11.17}{\db} for the \qty{10}{\centi\meter}, \qty{20.34}{\db} for \qty{30}{\centi\meter}, and \qty{22.95}{\db} for the \qty{50}{\centi\meter} distance for the wired scenario.
This proofs and quantifies the significant impact of classical \ac{DAQ} on measurements in the realm of capacitive \ac{HBC}.
\revise{The obtained channel gain for wireless \ac{DAQ} shows comparable performance as described in \cite{avlani_2020}. However, a quantitative comparison of this work to the \ac{SoA} remains challenging as differences in the test setup, like device size and environment, can noticeably influence the measured absolute value.
Using the same test setup minimizes the board's effects, as capacitive coupling to the environment dominates the received signal.}

To compare the communication power of capacitive \ac{HBC} with \ac{BLE}, the most prominent \ac{RF} communication method for wearables, we define energy efficiency as a function of the data rate (energy/bit). When taking the measured Tx power of \qty{2.71}{\milli\watt} at \qty{64}{\mega\hertz} and assuming a reasonable data rate of \qty{1}{\mbps} \cite{wi-R_white_paper}, we achieve \qty[per-mode = symbol]{2.71}{\nano\joule\per\bit}. This is comparable to Bluetooth, which achieves energy efficiencies larger than \qty[per-mode = symbol]{1}{\nano\joule\per\bit} up to \qty[per-mode = symbol]{15}{\nano\joule\per\bit} \cite{wi-R_datta_2023}.
With a simple Tx \revise{frontend}, capacitive \ac{HBC} is able to achieve comparable energy efficiency to \ac{BLE}.
\revise{With a Tx power of \qty{2.71}{\milli\watt} at \qty{64}{\mega\hertz}, the maximal received power for distances of \qty{10}{\centi\meter}, \qty{30}{\centi\meter}, and \qty{50}{\centi\meter} lies at \qty{-60}{\dbm}, \qty{-69.9}{\dbm}, and \qty{-71.9}{\dbm}, respectively. Although these power levels are very small, they appear to converge to a fixed value at larger distances. This makes them interesting for energy harvesting applications, especially if the baseline of the received power can be increased.}

\section{Conclusion}\label{sec:conclusion}
While conventional data acquisition setups are effective for quantifying the forward path loss, which depends on the conductive properties of the human body, they substantially alter the return path behavior by artificially modifying the capacitive coupling to earth ground.
Therefore, a wireless, wearable-sized data acquisition system is essential for quantitatively evaluating the full \ac{HBC} communication channel in a realistic environment with minimal measurement interference. 
To address this challenge, this work introduces \textit{BodySense}, an evaluation platform for human body communication that is fully wireless, compact enough for wearable applications, and designed for extendability.
To validate the proposed system, the measured channel gains of a classical, grid-connected setup and a wireless setup have been determined for distances of \qty{10}{\centi\meter}, \qty{30}{\centi\meter}, and \qty{50}{\centi\meter} between transmitter and receiver for a frequency range between \qty{4}{\mega\hertz} and \qty{64}{\mega\hertz}.
A comparison between the two scenarios yields an average overestimation of \qty{18.15}{\db} over all investigated distances for the classical case, highlighting the importance of evaluating capacitive \ac{HBC} in realistic conditions.
When comparing the energy consumption of capacitive \ac{HBC} with \ac{BLE}, we achieved results comparable to state-of-the-art \ac{BLE} frontends. 
This demonstrates its potential as a promising alternative to conventional \ac{RF} links, offering opportunities to further enhance the overall energy efficiency of wearable devices and move closer to the realization of battery-free, body-worn sensor nodes.


\begin{acronym}

    \acro{BAN}{Body Area Network}
    \acro{WBAN}{Wireless Body Area Network}
    \acro{PAN}{Personal Area Network}
    \acro{IoT}{Internet of Things}
    \acro{IoB}{Internet of Bodies}
    \acro{AI}{Artificial Intelligence}
    \acro{MCU}{Microcontroller}
    \acro{GNSS}{Global Navigation Satellite System}
    \acro{Nb-IoT}{Narrowband IoT}
    \acro{LoRa}{Long Range}
    \acro{LoRaWAN}{Long Range Wide Area Network}
    \acro{BLE}{Bluetooth Low Energy}
    \acro{UWB}{Ultra-Wideband}
    \acro{NFC}{near Field Communication}
    \acro{SpO2}{Oxigen Saturation}
    \acro{VR}{Virtual Reality}
    \acro{AR}{Augmented Reality}
    \acro{EQS-HBC}{Electro-Quasistatic Human Body Communication}
    \acro{EQS}{Electro-Quasistatic}
    \acro{HBC}{Human Body Communication}
    \acro{WBAN}{Wireless Body Area Network}
    \acro{WPAN}{Wireless Personal Area Network}
    \acro{WP}{Work Package}
    \acro{CSMA}{Carrier Sense Multiple Access}
    \acro{KB}{Kilobyte}
    \acro{PMIC}{Power Management IC}

    \acro{HCI}{Human-Computer Interaction}
    \acro{HMI}{Human-Machine Interaction}
    \acro{TIA}{Transimpedance Amplifier}
    \acro{SoTA}{State-of-The-Art}
    \acro{SoA}{State-of-Art}
    \acro{WLAN}{Wireless Local Area Network}
    \acro{PCE}{power Conversion Efficiency}
    \acro{ECG}{Electrocardiogram}
    \acro{BOM}{Bill of Material}
    \acro{DAQ}{Data Acquisition}
    \acro{RSSI}{Received Signal Strength Indicator}
    \acro{RSS}{Received Signal Strength}
    \acro{GBP}{Gaind-Bandwidth Product}

    \acro{LoS}{Line-of-Sight}
    \acro{nLoS}{non-Line-of-Sight}
    \acro{QoS}{Quality of Service}
    \acro{NB}{Narrowband Communication}
    \acro{WPT}{Wireless Power Transfer}
    \acro{IBPT}{Intra-Body Power Transfer}
    \acro{ICNIRP}{International Commission on Non-Ionizing Radiation Protection} 
    \acro{SAR}{Specific Energy Absorption}

    \acro{RF}{Radio Frequency}
    \acro{RFID}{Radio Frequency Identification}
    \acro{IoT}{Internet of Things}
    \acro{IoUT}{Internet of Underwater Things}
    \acro{UWN}{Underwater Wireless Network}
    \acro{UWSN}{Underwater Wireless Sensor Node}
    \acro{AUV}{Autonomous Underwater Vehicles}
    \acro{UAC}{Underwater Acoustic Channel}
    \acro{FSK}{Frequency Shift Keying}
    \acro{OOK}{On-Off Keying}
    \acro{ASK}{Amplitude Shift Keying}
    \acro{UUID}{Universal Unique Identifier}
    \acro{PZT}{Lead Zirconium Titanate}
    \acro{AC}{Alternating Current}
    \acro{NVC}{Negative Voltage Converter}
    \acro{NVCR}{Negative Voltage Converter Rectifier}
    \acro{FWR}{Full-Wave Rectifier}
    \acro{GPIO}{General Purpose Input/Output}
    \acro{PCB}{Printed Circuit Board}
    \acro{AUV}{Autonomous Underwater Vehicle}
    \acro{IMU}{Intertial Measurement Unit}
    \acro{BLE}{Bluetooth Low Energy}
    \acro{FSR}{Force Sensing Resistor}
    \acro{SiP}{System in Package}
    \acro{SoC}{System on Chip}
    \acro{SpO2}{Oxigen Saturation}
    \acro{PULP}{Parallel Ultra-Low Power}
    \acro{ML}{Machine Learning}
    \acro{ADC}{Analog to Digital Converter}
    \acro{TCDM}{Tightly Coupled Data Memory}
    \acro{GNSS}{Global Navigation Satellite System}
    \acro{IC}{Integrated Circuit}
    \acro{POM}{Polyoxymethylene}
    \acro{RTOS}{Real-Time Operating System}
    \acro{LoRaWAN}{Long Range Wide Area Network}
    \acro{ML}{Machine Learning}
    \acro{IIR}{Infinite Impulse Response}
    \acro{PLL}{Phase-locked Loop}
\end{acronym}

\section*{ACKNOWLEDGMENT}
This work was found by the Swiss National Science Foundation SNSF under the projects “BodyLink: Enabling Battery-free body-worn Sensing and Communication with Energy Transfer” (Grant Nr. 220867) and “Wearable Nano-Opto-electro-mechanic Systems” (Grant Nr. 209675).
\bibliographystyle{IEEEtranDOI} 


\end{document}